\documentclass[12pt,letterpaper,english,aps,superscriptaddress,groupedaddress,showpacs,showkeys]{revtex4}
\usepackage[T1]{fontenc}
\usepackage[latin9]{inputenc}
\usepackage{babel}
\usepackage{amsmath}
\usepackage{amssymb}
\usepackage{esint}
\usepackage[unicode=true]
 {hyperref}
\usepackage{breakurl}

\makeatletter

\@ifundefined{textcolor}{}
{%
 \definecolor{BLACK}{gray}{0}
 \definecolor{WHITE}{gray}{1}
 \definecolor{RED}{rgb}{1,0,0}
 \definecolor{GREEN}{rgb}{0,1,0}
 \definecolor{BLUE}{rgb}{0,0,1}
 \definecolor{CYAN}{cmyk}{1,0,0,0}
 \definecolor{MAGENTA}{cmyk}{0,1,0,0}
 \definecolor{YELLOW}{cmyk}{0,0,1,0}
}

\usepackage{dcolumn}% Align table columns on decimal point
\usepackage{bm}% bold math
\usepackage{amsfonts}% note how statements can be commented out

\makeatother

\begin{document}

\title{On the boundary conditions for the 1D Weyl-Majorana particle in a
box}

\author{Salvatore De Vincenzo}

\homepage{https://orcid.org/0000-0002-5009-053X}

\email{[salvatore.devincenzo@ucv.ve]}

\selectlanguage{english}%

\affiliation{Escuela de F\'{\i}sica, Facultad de Ciencias, Universidad Central de Venezuela,
A.P. 47145, Caracas 1041-A, Venezuela.}

\thanks{I would like to dedicate this article to the memory of Luis A. Gonz\'{a}lez-D\'{\i}az,
friend and physicist.}

\date{September 29, 2020}
\begin{abstract}
\noindent \textbf{Abstract} In (1+1) space-time dimensions, we can
have two particles that are Weyl and Majorana particles at the same
time---1D Weyl-Majorana particles. That is, the right-chiral and left-chiral
parts of the two-component Dirac wave function that satisfies the
Majorana condition, in the Weyl representation, describe these particles,
and each satisfies their own Majorana condition. Naturally, the nonzero
component of each of these two two-component wave functions satisfies
a Weyl equation. We investigate and discuss this issue and demonstrate
that for a 1D Weyl-Majorana particle in a box, the nonzero components,
and therefore the chiral wave functions, only admit the periodic and
antiperiodic boundary conditions. From the latter two boundary conditions,
we can only construct four boundary conditions for the entire Dirac
wave function. Then, we demonstrate that these four boundary conditions
are also included within the most general set of self-adjoint boundary
conditions for a 1D Majorana particle in a box.
\end{abstract}



\maketitle

%%%%%%%
\section{Introduction}
%%%%%%%

\noindent The equation for a first quantized free massless Dirac single
particle in (1+1) dimensions---or the one-dimensional free massless
Dirac particle---has the form
\begin{equation}
\mathrm{i}\hat{\gamma}^{\mu}\partial_{\mu}\Psi=0,
\end{equation}
where $\Psi=\Psi(x,t)$ is a two-component wave function---a Dirac
wave function, $\partial_{\mu}=(c^{-1}\partial_{t},\partial_{x})$
(as usual), and the Dirac matrices $\hat{\gamma}^{\mu}$, with $\mu=0,1$,
satisfy the relations $\hat{\gamma}^{\mu}\hat{\gamma}^{\nu}+\hat{\gamma}^{\nu}\hat{\gamma}^{\mu}=2g^{\mu\nu}\hat{1}_{2}$,
where $g^{\mu\nu}=\mathrm{diag}(1,-1)$ ($\hat{1}_{2}$ is the $2\times2$
identity matrix), and $(\hat{\gamma}^{\mu})^{\dagger}=\hat{\gamma}^{0}\hat{\gamma}^{\mu}\hat{\gamma}^{0}$
(the symbol $^{\dagger}$ denotes the Hermitian conjugate, or the
adjoint, of a matrix and an operator) \cite{RefA}. 

The so-called charge-conjugate wave function, $\Psi_{C}\equiv\hat{S}_{C}\Psi^{*}$,
also satisfies Eq. (1), namely, 
\begin{equation}
\mathrm{i}\hat{\gamma}^{\mu}\partial_{\mu}\Psi_{C}=0,
\end{equation}
but this implies that 
\begin{equation}
\hat{S}_{C}\,(\mathrm{i}\hat{\gamma}^{\mu})^{*}\hat{S}_{C}^{-1}=\mathrm{i}\hat{\gamma}^{\mu},
\end{equation}
where $\hat{S}_{C}$ is the charge-conjugation matrix (the superscript
asterisk$^{*}$ represents the complex conjugate) \cite{RefB,RefC}.
The latter matrix can be chosen to be unitary (up to a phase factor)
\cite{RefA}.

Let us introduce the following wave functions:
\begin{equation}
\Psi_{\pm}\equiv\frac{1}{2}\left(\hat{1}_{2}\pm\hat{\Gamma}^{5}\right)\Psi,
\end{equation}
where the (Hermitian) matrix $\hat{\Gamma}^{5}\equiv\hat{\gamma}^{0}\hat{\gamma}^{1}$
is the chirality matrix, and satisfies the relations $(\hat{\Gamma}^{5})^{2}=\hat{1}_{2}$
and $\hat{\Gamma}^{5}\hat{\gamma}^{\mu}+\hat{\gamma}^{\mu}\hat{\Gamma}^{5}=\hat{0}_{2}$
($\hat{0}_{2}$ is the $2$-dimensional zero matrix) \cite{RefD,RefE}.
In addition, $\hat{\Gamma}^{5}$ satisfies the relation $\hat{S}_{C}\,(\mathrm{i}\hat{\Gamma}^{5})^{*}(\hat{S}_{C})^{-1}=-\mathrm{i}\hat{\Gamma}^{5}$,
and 
\begin{equation}
\left[\frac{1}{2}\left(\hat{1}_{2}\pm\hat{\Gamma}^{5}\right)\right]^{2}=\frac{1}{2}\left(\hat{1}_{2}\pm\hat{\Gamma}^{5}\right)\,,\quad\mathrm{and}\quad\frac{1}{2}\left(\hat{1}_{2}\pm\hat{\Gamma}^{5}\right)\frac{1}{2}\left(\hat{1}_{2}\mp\hat{\Gamma}^{5}\right)=\hat{0}_{2}.
\end{equation}
Note that, the two-component Dirac wave functions $\Psi_{+}$ (which
must also satisfy the relations $\frac{1}{2}(\hat{1}_{2}+\hat{\Gamma}^{5})\Psi_{+}=\Psi_{+}$
and $\frac{1}{2}(\hat{1}_{2}-\hat{\Gamma}^{5})\Psi_{+}=0$) and $\Psi_{-}$
(which must also satisfy the relations $\frac{1}{2}(\hat{1}_{2}-\hat{\Gamma}^{5})\Psi_{-}=\Psi_{-}$
and $\frac{1}{2}(\hat{1}_{2}+\hat{\Gamma}^{5})\Psi_{-}=0$) are eigenstates
of $\hat{\Gamma}^{5}$. $\Psi_{+}$ is called the right-chiral eigenstate
(eigenvalue $+1$) and $\Psi_{-}$ the left-chiral eigenstate (eigenvalue
$-1$). The charge conjugate of the wave functions $\Psi_{\pm}$ verify
that $(\Psi_{\pm})_{C}=(\Psi_{C})_{\pm}$, i.e., both $\Psi_{+}$
and $(\Psi_{+})_{C}$ are right-chiral states, and similarly, both
$\Psi_{-}$ and $(\Psi_{-})_{C}$ are left-chiral states. This is
not the case in (3+1) dimensions \cite{RefE}.

Now, note that by multiplying the Dirac equation in Eq. (1) by $\tfrac{1}{2}(\hat{1}_{2}+\hat{\Gamma}^{5})$
from the left, we obtain the equation
\begin{equation}
\mathrm{i}\hat{\gamma}^{\mu}\partial_{\mu}\Psi_{-}=0,
\end{equation}
and similarly, multiplying Eq. (1) by $\tfrac{1}{2}(\hat{1}_{2}-\hat{\Gamma}^{5})$,
we obtain
\begin{equation}
\mathrm{i}\hat{\gamma}^{\mu}\partial_{\mu}\Psi_{+}=0.
\end{equation}
Certainly, because $\Psi=\Psi_{+}+\Psi_{-}$, the latter pair of equations
is equivalent to the Dirac equation. Because the charge-conjugate
wave function $\Psi_{C}$ also satisfies the Dirac equation, we also
have two equations equivalent to the latter equation. Specifically,
by multiplying the Dirac equation for $\Psi_{C}$ by $\tfrac{1}{2}(\hat{1}_{2}+\hat{\Gamma}^{5})$
and $\tfrac{1}{2}(\hat{1}_{2}-\hat{\Gamma}^{5})$, we obtain
\begin{equation}
\mathrm{i}\hat{\gamma}^{\mu}\partial_{\mu}(\Psi_{-})_{C}=0\,,\quad\mathrm{and}\quad\mathrm{i}\hat{\gamma}^{\mu}\partial_{\mu}(\Psi_{+})_{C}=0,
\end{equation}
respectively (remember that $(\Psi_{\pm})_{C}=\hat{S}_{C}\Psi_{\pm}^{*}$).
So far, $\Psi_{-}$ and $(\Psi_{-})_{C}$, and $\Psi_{+}$ and $(\Psi_{+})_{C}$,
are different wave functions, but they (all) satisfy the same two-component
equation of motion.

The so-called (Lorentz-covariant) Majorana condition (see, for example,
Refs. \cite{RefE,RefF}),
\begin{equation}
\Psi=\Psi_{C},
\end{equation}
imposed upon the two-component Dirac wave function $\Psi$ gives us
the following relations:
\begin{equation}
\Psi_{+}=(\Psi_{+})_{C}\quad\mathrm{and}\quad\Psi_{-}=(\Psi_{-})_{C}.
\end{equation}
Thus, in (1+1) dimensions, if $\Psi$ satisfies the Majorana condition,
then both $\Psi_{+}$ and $\Psi_{-}$ satisfy this condition. Clearly,
the pair of Eqs. (6) and (7) and the pair of restrictions in (10)
describe a massless Majorana particle in (1+1) dimensions. Naturally,
by imposing the Majorana condition on the equations in (8), we again
obtain Eqs. (6) and (7). 

In the Weyl representation, the two-component wave function and the
Dirac matrices can be written as follows \cite{RefE}: 
\begin{equation}
\Psi\equiv\left[\begin{array}{c}
\varphi_{1}\\
\varphi_{2}
\end{array}\right]\,,\quad\hat{\gamma}^{0}=\hat{\sigma}_{x}\,,\quad\hat{\gamma}^{1}=-\mathrm{i}\hat{\sigma}_{y}
\end{equation}
($\hat{\sigma}_{x}$ and $\hat{\sigma}_{y}$ are Pauli matrices).
By substituting $\Psi$, $\hat{\gamma}^{0}$ and $\hat{\gamma}^{1}$
from Eq. (11) into Eq. (1), we obtain two decoupled differential equations,
namely,
\begin{equation}
\begin{array}{c}
\mathrm{i}\hbar\left(\partial_{t}+c\,\partial_{x}\right)\varphi_{1}=0,\\
\mathrm{i}\hbar\left(\partial_{t}-c\,\partial_{x}\right)\varphi_{2}=0.
\end{array}
\end{equation}
Likewise, in the Weyl representation, we have that $\hat{\Gamma}^{5}=\hat{\sigma}_{z}$,
$\Psi_{+}=\frac{1}{2}(\hat{1}_{2}+\hat{\Gamma}^{5})\Psi=\left[\,\varphi_{1}\;0\,\right]^{\mathrm{T}}$,
and $\Psi_{-}=\frac{1}{2}(\hat{1}_{2}-\hat{\Gamma}^{5})\Psi=\left[\,0\;\varphi_{2}\,\right]^{\mathrm{T}}$,
as expected ($^{\mathrm{T}}$ represents the transpose of a matrix);
thus, the first of the equations in (12) can also be obtained from
Eq. (7), and the second equation can be obtained from Eq. (6) (also
as expected). Naturally, the wave function $\Psi$ that describes
the one-dimensional Dirac particle has two independent complex components,
or two complex degrees of freedom, i.e., four real degrees of freedom.
On the other hand, as said before, the Dirac equation (1) can also
describe a one-dimensional massless Majorana single particle if, in
addition, $\Psi$ complies with the Majorana condition, $\Psi=\Psi_{C}$.
Again, in the Weyl representation, we can write 
\begin{equation}
\hat{S}_{C}=\exp(\mathrm{i}\nu)\hat{\gamma}^{0}\hat{\gamma}^{1}=\exp(\mathrm{i}\nu)\hat{\sigma}_{z}\,(\propto\hat{\Gamma}^{5}),
\end{equation}
where $\nu=[0,2\pi)$ is an arbitrary phase certainly not fixed by
Eq. (3) but by us. Consequently, the Majorana condition in the form
given in Eq. (10) gives us the following two (independent) relations:
\begin{equation}
\varphi_{1}=\exp(\mathrm{i}\nu)\,\varphi_{1}^{*}\quad\mathrm{and}\quad\varphi_{2}=-\exp(\mathrm{i}\nu)\,\varphi_{2}^{*}.
\end{equation}
Thus, the equation that describes the massless Majorana particle in
(1+1) dimensions, in the Weyl representation, is a pair of decoupled
one-component first-order equations, i.e., the pair of equations in
(12), with the restrictions given in Eq. (14). In the end, the wave
function $\Psi$ that describes this kind of particle only has two
real degrees of freedom (half of those of the Dirac particle, but
it has the same degrees of freedom as that of the Weyl particle, as
we will see below).

At this point, certain remarks are in order. The wave function $\Psi$
for the one-dimensional massless Majorana particle must satisfy the
one-dimensional Dirac equation (in the so-called chiral limit $\mathrm{m}=0$,
or the massless limit) as well as the Majorana condition (thus, ultimately,
we could also call it a massless (Dirac)-Majorana particle). This
is true in any representation. Precisely, the most general set of
boundary conditions for this particle when it is within a box, in
the Weyl representation, was presented in Ref. \cite{RefE}. {[}In
fact, the latter reference dealt with the massive Majorana particle,
but the most general set of boundary conditions presented there does
not depend on the value of the mass.{]} This set consists of two one-parametric
families of (complex) boundary conditions for the Dirac wave function
$\Psi=\left[\,\varphi_{1}\;\varphi_{2}\,\right]^{\mathrm{T}}$ (we
write and use them in section II). Certainly, all these boundary conditions
arise when the self-adjointness condition is imposed on the Dirac
Hamiltonian operator of the system, namely, $\hat{\mathrm{H}}=-\mathrm{i}\hbar c\,\hat{\sigma}_{z}\partial_{x}=\hat{\mathrm{H}}^{\dagger}$
(remember that the Dirac equation in Eq. (1) in its canonical form
is $\mathrm{i}\hbar\,\partial_{t}\Psi=\hat{\mathrm{H}}\Psi$), and
inside its domain $\mathcal{D}(\hat{\mathrm{H}})=\mathcal{D}(\hat{\mathrm{H}}^{\dagger})$,
we have precisely only these boundary conditions. Incidentally, in
Ref. \cite{RefE}, $\nu=3\pi/2$ was specifically selected when choosing
the charge-conjugation matrix $\hat{S}_{C}$ (see Eq. (13)); nevertheless,
this choice does not change the two families of boundary conditions.

Let us now return to the pair of equations in (12), forgetting how
we obtained them. Precisely, these equations would be the (free) Weyl
equations in (1+1) dimensions \cite{RefG}. Each of the equations
in (12) would describe a specific type of one-dimensional uncharged
Weyl particle (a Weyl particle is always massless). In fact, $\varphi_{1}$
and $\varphi_{2}$ in Eq. (12) are transformed in two different ways
under the Lorentz boost, i.e., they transform according to two inequivalent
representations of the Lorentz group \cite{RefE,RefH}. Now, note
that, if, for example, the one-component wave function $\varphi_{1}$
that describes this kind of particle is complex-valued, then $\varphi_{1}$
has one complex degree of freedom, i.e., two real degrees of freedom.
The same is valid for the one-component wave function $\varphi_{2}$.
Thus, a one-dimensional Weyl particle has the same degrees of freedom
as the one-dimensional massless Majorana particle.

It can be noticed, although not without some surprise, that from the
results given in Eqs. (6) and (7), as well as (10), one can introduce
two other types of (relativistic) one-dimensional particles. Indeed,
the equation that describes the first type of one-dimensional particle
is given by Eq. (6), with its respective restriction given in Eq.
(10), namely,
\begin{equation}
\mathrm{i}\hat{\gamma}^{\mu}\partial_{\mu}\Psi_{-}=0\:\,\mathrm{and}\:\,\Psi_{-}=(\Psi_{-})_{C}.
\end{equation}
Similarly, the equation that describes the second type of one-dimensional
particle is given by Eq. (7) with its respective restriction given
in Eq. (10), namely,
\begin{equation}
\mathrm{i}\hat{\gamma}^{\mu}\partial_{\mu}\Psi_{+}=0\:\,\mathrm{and}\:\,\Psi_{+}=(\Psi_{+})_{C}.
\end{equation}
Clearly, the two-component Dirac wave functions with definite chirality,
$\Psi_{+}$ and $\Psi_{-}$, each satisfy the one-dimensional Dirac
equation (in the massless limit) and their own Majorana conditions
(thus, in principle, we could call these particles Dirac-Majorana
particles again). However, it is only in the Weyl representation that
it is explicitly shown that $\Psi_{+}$ and $\Psi_{-}$ have each
one-only one-nonzero (complex) components, which are transformed independently
under the Lorentz transformation (or the Lorentz boost). Certainly,
this Lorentz transformation does not change the chirality of the wave
function \cite{RefH}. We mention in passing that due to their characteristics,
the wave functions $\Psi_{+}$ and $\Psi_{-}$ are sometimes also
called Weyl wave functions or said to satisfy the Weyl condition (thus,
certainly, we could call the particles described by these wave functions
Weyl-Majorana particles) \cite{RefI}. The possibility that a wave
function in (1+1) dimensions (and in other distinct space-time dimensions)
can simultaneously satisfy the aforementioned Weyl condition (in even
dimensions) and that of Majorana has been noted in the literature.
For more details on this issue, see, for example, Ref. \cite{RefI},
appendix B, and Ref. \cite{RefJ}, pp. 35-45. 

Thus, from the results in (15), we can say that the first type of
one-dimensional particle is completely defined by
\begin{equation}
\mathrm{i}\hbar\left(\partial_{t}-c\,\partial_{x}\right)\varphi_{2}=0\,,\quad\mathrm{with}\quad\varphi_{2}=-\exp(\mathrm{i}\nu)\,\varphi_{2}^{*}
\end{equation}
(if $\varphi_{2}\in\mathbb{C}$, then we just have here one real degree
of freedom). Similarly, from the results in (16), we can say that
the second type of one-dimensional particle is completely defined
by
\begin{equation}
\mathrm{i}\hbar\left(\partial_{t}+c\,\partial_{x}\right)\varphi_{1}=0\,,\quad\mathrm{with}\quad\varphi_{1}=\exp(\mathrm{i}\nu)\,\varphi_{1}^{*}
\end{equation}
(and again, if $\varphi_{1}\in\mathbb{C}$, then we just have one
real degree of freedom). Clearly, the one-component (Weyl) wave functions
$\varphi_{1}$ and $\varphi_{2}$ satisfy each a one-dimensional Weyl
equation and a condition that comes from imposing the Majorana condition
on the entire wave function $\Psi=\left[\,\varphi_{1}\;\varphi_{2}\,\right]^{\mathrm{T}}$,
i.e., on the wave functions $\Psi_{+}=\left[\,\varphi_{1}\;0\,\right]^{\mathrm{T}}$
and $\Psi_{-}=\left[\,0\;\varphi_{2}\,\right]^{\mathrm{T}}$. In this
way, we can now decide to call these particles Weyl-Majorana particles,
i.e., two particles that are each a Weyl particle and a Majorana particle
at the same time. Certainly, each Weyl-Majorana particle has half
the (real) degrees of freedom of the Weyl particle as well as the
Majorana particle. In the next section, we find the physically acceptable
boundary conditions for this type of particle when it can only be
inside a box.

%%%%%%%
\section{A 1D Weyl-Majorana particle in a box}
%%%%%%%

\noindent Let us consider a one-dimensional Weyl-Majorana particle
in a box of size $L$, with ends, for example, at $x=0$ and $x=L$.
First, we write the two Weyl equations in Eqs. (17) and (18) in their
canonical forms in a single equation as follows:
\begin{equation}
\mathrm{i}\hbar\,\partial_{t}\varphi_{a}=\hat{\mathrm{h}}_{a}\varphi_{a}\,,\quad a=1,2,
\end{equation}
where
\begin{equation}
\hat{\mathrm{h}}_{a}\equiv-\mathrm{i}\hbar c(-1)^{a-1}\partial_{x}
\end{equation}
is the formally self-adjoint, or Hermitian, one-dimensional Weyl Hamiltonian
operator, i.e., $\hat{\mathrm{h}}_{a}=\hat{\mathrm{h}}_{a}^{\dagger}$
(i.e., essentially without the specification of its domain). Clearly,
$\hat{\mathrm{h}}_{a}$ is very similar to the usual nonrelativistic
momentum operator (see, for example, Ref. \cite{RefK}). The Hamiltonian
$\hat{\mathrm{h}}_{a}$ is also a self-adjoint operator; this is essentially
because its domain, i.e., the set of Weyl one-component wave functions
$\varphi_{a}=\varphi_{a}(x,t)$ in the Hilbert space of the square
integrable functions $\mathcal{H}=\mathcal{L}^{2}[0,L]$ on which
$\hat{\mathrm{h}}_{a}$ can act ($\equiv\mathcal{D}(\hat{\mathrm{h}}_{a})\subset\mathcal{H}$),
includes the following general boundary condition dependent on a single
parameter, namely,
\begin{equation}
\varphi_{a}(L,t)=\exp(\mathrm{i}\theta)\,\varphi_{a}(0,t),
\end{equation}
with $\theta\in[0,2\pi)$; in addition, $\hat{\mathrm{h}}_{a}\varphi_{a}\in\mathcal{H}$
\cite{RefK}. Moreover, the scalar product in $\mathcal{H}$ is denoted
by $\langle\psi_{a},\chi_{a}\rangle\equiv\int_{0}^{L}\,\mathrm{d}x\,\psi_{a}^{*}\chi_{a}$,
and the norm is $\left\Vert \,\psi_{a}\,\right\Vert \equiv\sqrt{\langle\psi_{a},\psi_{a}\rangle}$.
Precisely, $\hat{\mathrm{h}}_{a}$ satisfies the hermiticity condition,
or the self-adjointness condition, namely,
\begin{equation}
\langle\psi_{a},\hat{\mathrm{h}}_{a}\chi_{a}\rangle=\langle\hat{\mathrm{h}}_{a}\psi_{a},\chi_{a}\rangle-\left.\mathrm{i}\hbar c(-1)^{a-1}\left[\,\psi_{a}^{*}\chi_{a}\,\right]\right|_{0}^{L}=\langle\hat{\mathrm{h}}_{a}\psi_{a},\chi_{a}\rangle,
\end{equation}
where we introduce the notation $\left.\left[\, f\,\right]\right|_{0}^{L}\equiv f(x=L,t)-f(x=0,t)$,
and $\psi_{a}$ and $\chi_{a}$ are Weyl wave functions in $\mathcal{D}(\hat{\mathrm{h}}_{a})=\mathcal{D}(\hat{\mathrm{h}}_{a}^{\dagger})$.
Now, note that the Majorana condition imposed on the Weyl wave function
$\varphi_{a}$, namely, $\varphi_{a}=(-1)^{a-1}\exp(\mathrm{i}\nu)\,\varphi_{a}^{*}$
(see Eqs. (17) and (18)), implies that $\varphi_{a}^{*}$ must also
comply the general boundary condition in Eq. (21), in which case the
phase $\exp(\mathrm{i}\theta)$ in Eq. (21) must be real. Therefore,
$\theta=0,\pi$; thus, the boundary conditions for a one-dimensional
free Weyl-Majorana particle in a box can only be the periodic boundary
condition, $\varphi_{a}(L,t)=\varphi_{a}(0,t)$, and the antiperiodic
boundary condition, $\varphi_{a}(L,t)=-\varphi_{a}(0,t)$. Incidentally,
these two boundary conditions are nonconfining boundary conditions,
i.e., neither of these can cancel the probability current density
at the ends of the box. In effect, in this case, the probability current
density (corresponding to the wave function $\varphi_{a}$) is given
by $j_{a}\equiv(-1)^{a-1}c\,\varphi_{a}^{*}\varphi_{a}$, where $(\varphi_{a}^{*}\varphi_{a})(x=L,t)=(\varphi_{a}^{*}\varphi_{a})(x=0,t)$
(the latter relation comes out of Eq. (22) and must be satisfied by
all boundary conditions in the domain of $\hat{\mathrm{h}}_{a}$)
\cite{RefH}. Thus, we now also have the relation $j_{a}(x=L,t)=j_{a}(x=0,t)$,
which is obviously satisfied by the periodic and antiperiodic boundary
conditions and by all boundary conditions within Eq. (21). We mention
in passing that, because the solutions of the Weyl equations in (19)
can always be chosen to be real, these solutions could only admit
boundary conditions that are within Eq. (21) with the condition $\varphi_{a}=\varphi_{a}^{*}$,
which implies that only periodic and antiperiodic boundary conditions
could be imposed. The latter point was recently noted in Ref. \cite{RefH}. 

Thus, for the Weyl-Majorana particle described by the two-component
wave function $\Psi_{-}=\left[\,0\;\varphi_{2}\,\right]^{\mathrm{T}}$,
which satisfies the results in (15), the two boundary conditions given
above (for $\varphi_{2}$) must be written as follows (we omit the variable $t$
in the boundary conditions hereinafter):
\begin{equation}
\Psi_{-}(L)=\Psi_{-}(0)
\end{equation}
(the periodic boundary condition), and 
\begin{equation}
\Psi_{-}(L)=-\Psi_{-}(0)
\end{equation}
(the antiperiodic boundary condition). Similarly, for the Weyl-Majorana
particle described by the two-component wave function $\Psi_{+}=\left[\,\varphi_{1}\;0\,\right]^{\mathrm{T}}$,
which satisfies the results in (16), the boundary conditions for $\varphi_{1}$
must now be written as follows:
\begin{equation}
\Psi_{+}(L)=\Psi_{+}(0),
\end{equation}
and 
\begin{equation}
\Psi_{+}(L)=-\Psi_{+}(0)
\end{equation}
(again, the periodic and antiperiodic boundary conditions). These
are the only boundary conditions that can be imposed on $\Psi_{-}$
and $\Psi_{+}$ when they describe a one-dimensional Weyl-Majorana
particle. 

From the boundary conditions in Eqs. (23)-(26), just four boundary
conditions for the Dirac wave function $\Psi$ can be constructed.
In effect, because $\Psi=\Psi_{+}+\Psi_{-}$, where $\Psi_{+}$ and
$\Psi_{-}$ are given in Eq. (4), the following results are obtained:
\begin{equation}
\Psi(L)=-\hat{\Gamma}^{5}\Psi(0),
\end{equation}
which comes from the conditions in Eqs. (23) and (26), and 
\begin{equation}
\Psi(L)=\hat{\Gamma}^{5}\Psi(0),
\end{equation}
which comes from the conditions in Eqs. (24) and (25). Likewise, 
\begin{equation}
\Psi(L)=\Psi(0),
\end{equation}
which comes from the conditions in Eqs. (23) and (25), and 
\begin{equation}
\Psi(L)=-\Psi(0),
\end{equation}
which comes from the conditions in Eqs. (24) and (26). On the other
hand, it can be shown (as we do below) that these four boundary conditions
are in fact included within the most general set of self-adjoint boundary
conditions for the one-dimensional (either massive or massless) Majorana
particle enclosed in a box. In effect, this set is formed by the following
two families of boundary conditions for the Dirac wave function $\Psi=\left[\,\varphi_{1}\;\varphi_{2}\,\right]^{\mathrm{T}}$,
in the Weyl representation (see Eqs. (35) and (36) in Ref. \cite{RefE}):
\begin{equation}
\Psi(L)=\frac{1}{m_{2}}\left[\begin{array}{cc}
-1 & -\mathrm{i}m_{0}\\
-\mathrm{i}m_{0} & +1
\end{array}\right]\Psi(0),
\end{equation}
where $(m_{0})^{2}+(m_{2})^{2}=1$, and
\begin{equation}
\Psi(L)=\frac{1}{m_{1}}\left[\begin{array}{cc}
+1 & -\mathrm{i}m_{3}\\
+\mathrm{i}m_{3} & +1
\end{array}\right]\Psi(0),
\end{equation}
where $(m_{1})^{2}+(m_{3})^{2}=1$; in addition, $m_{0}$, $m_{2}$
and $m_{1}$, $m_{3}$ are real quantities (more details of boundary
conditions for the problem of a Majorana particle in a box can also
be found in Refs. \cite{RefL} and \cite{RefM}). Then, in the first
subfamily above, one first notices that by setting $m_{0}=0$, one
has that $(m_{2})^{2}=1$, and therefore, $m_{2}=\pm1$. Thus, setting
$m_{0}=0$ and $m_{2}=+1$ in Eq. (31), one obtains the boundary condition
in Eq. (27), and setting $m_{0}=0$ and $m_{2}=-1$ in Eq. (31), one
obtains the boundary condition in Eq. (28). Likewise, in the second
subfamily above, one first notices that by setting $m_{3}=0$, we
have that $(m_{1})^{2}=1$, and therefore $m_{1}=\pm1$. Thus, setting
$m_{3}=0$ and $m_{1}=+1$ in Eq. (32), one obtains the boundary condition
in Eq. (29), and setting $m_{3}=0$ and $m_{1}=-1$ in Eq. (32), one
obtains the boundary condition in Eq. (30). Thus, the boundary conditions
in Eqs. (27)-(30) are also included within the most general set of
self-adjoint boundary conditions for the 1D Majorana particle in a
box. Because in this case the boundary conditions for $\Psi$ are
obtained from the physically acceptable boundary conditions for $\Psi_{+}$
and $\Psi_{-}$, we can say that $\Psi=\Psi_{+}+\Psi_{-}$ nevertheless
continues to describe a massless one-dimensional Majorana particle
in a box. 

%%%%%%%
\section{Conclusions}
%%%%%%%

\noindent Although a Majorana particle is generally considered a massive
particle and a Weyl particle is always massless, in (1+1) dimensions,
we can have particles that are Weyl and Majorana particles at the
same time; i.e., they are 1D Weyl-Majorana particles. Thus, the upper
and lower one-component wave functions of the two-component Dirac
wave function $\Psi$, in the Weyl representation, $\varphi_{1}$
and $\varphi_{2}$, each satisfy their own Weyl equation, and as we
know, each belongs to a different representation of the corresponding
Lorentz group. In addition, these wave functions are independent of
each other, i.e., they are not related after imposing the Majorana
condition on $\Psi=\left[\,\varphi_{1}\;\varphi_{2}\,\right]^{\mathrm{T}}$,
or equivalently, on the two-component Dirac wave functions with definite
chirality (or the chiral wave functions) $\Psi_{+}=\left[\,\varphi_{1}\;0\,\right]^{\mathrm{T}}$
and $\Psi_{-}=\left[\,0\;\varphi_{2}\,\right]^{\mathrm{T}}$. In constrast,
in (3+1) dimensions, the upper and lower two-component wave functions
of the four-component Dirac wave function (or bispinor), in the Weyl
representation, are linked by the Majorana condition \cite{RefE}.
Naturally, also in this case, these two two-component wave functions
each satisfy their own Weyl equation (certainly, in the massless limit
of the (free) Dirac equation), and each belongs to a different representation
of the corresponding Lorentz group.

To recap, in (1+1) dimensions, $\Psi_{+}$ and $\Psi_{-}$ each satisfy
the one-dimensional Dirac equation (in the massless limit) and their
own Majorana condition, but only in the Weyl representation are the
nonzero components of $\Psi_{+}$ and $\Psi_{-}$ transformed independently
under the Lorentz transformation; i.e., just under this circumstance,
$\Psi_{+}$ and $\Psi_{-}$ each describe a 1D Weyl-Majorana particle.

For a 1D Weyl-Majorana particle in a box, the chiral wave functions
only admit the periodic and antiperiodic boundary conditions. This
is because the one-component Weyl wave functions $\varphi_{1}$ and
$\varphi_{2}$ only admit these two boundary conditions. From these
two boundary conditions for $\Psi_{+}$ and $\Psi_{-}$, just four
boundary conditions can be constructed for the entire Dirac wave function
$\Psi=\Psi_{+}+\Psi_{-}$. Moreover, these four boundary conditions
are included within the most general set of self-adjoint boundary
conditions for a 1D (either massive or massless) Majorana particle
in a box. Thus, although these boundary conditions for $\Psi$ are
obtained from the boundary conditions for $\Psi_{+}$ and $\Psi_{-}$,
$\Psi=\Psi_{+}+\Psi_{-}$ continues to describe a 1D massless Majorana
particle in a box. We believe that our paper will be of interest to
all who are interested in the relativistic quantum mechanics in (1+1)
dimensions.

\begin{acknowledgments}
\noindent I thank Valedith Cusati, my wife, for all her support.
\end{acknowledgments}

\end{document}